\documentclass[cameraready]{Interspeech}

\title{Don't Listen to Me: A Lightweight, Low-Latency Model for Own-Voice Cancellation in Far-Field Speech Enhancement}

\author[affiliation={1},    orcid=0009-0005-9008-8708, correspondingauthor]{Mads}{Østergaard}
\author[affiliation={1},    orcid=0000-0001-6881-9766]{Alexander Neergaard}{Zahid}
\author[affiliation={1},    orcid=0009-0001-4448-3513]{Karl}{Ulbæk}
\author[affiliation={1,2},  orcid=0009-0004-5652-5372]{Andreas Hansen}{Bagge}
\author[affiliation={1,2},  orcid=0009-0007-5928-7279]{Kenny Falkjær}{Olsen}
\author[affiliation={3},    orcid=0000-0002-1851-7396]{Rasmus Malik Høegh}{Lindrup}

\address{
    $^1$ WS Audiology, Lynge, Denmark \\
    $^2$ DTU Compute, Technical University of Denmark, Kgs. Lyngby, Denmark \\
    $^3$ Verth, Denmark
}

\email{mads.oestergaard@wsa.com, alexander.neergaardzahid@wsa.com, karl.ulbaek@wsa.com, kenny.olsen@wsa.com, rasmus@verth.ai}


\usepackage[T1]{fontenc}
\usepackage{comment}
\usepackage{amsmath}
\usepackage{graphicx}
\usepackage{hyperref}
\usepackage{xcolor}
\usepackage{booktabs}
\usepackage{cleveref}
\crefname{figure}{Figure}{Figures}
\Crefname{figure}{Figure}{Figures}
\crefname{table}{Table}{Tables}
\Crefname{table}{Table}{Tables}
\crefname{equation}{Eq.}{Eqs.}
\crefname{equation}{Eq.}{Eqs.}
\usepackage{amssymb}
\usepackage{multirow}

\usepackage[skip=0.5\baselineskip]{caption}
\usepackage{enumitem}

\begin{document}

\maketitle
\begin{abstract}
  We introduce own-voice cancellation (OVC): removing a target (enrolled) speaker from a
  noisy multi-speaker mixture while preserving any remaining speech. Framed as the
  complement of target speaker extraction, OVC addresses latency-induced own-voice
  artifacts that arise when a far-field device streams enhanced audio back to the user,
  as the round-trip time easily exceeds the perceptual threshold for own-voice
  distortion. We condition a time-domain model with only 2\,ms algorithmic latency on a
  short enrollment utterance and benchmark TD-SpeakerBeam alongside a lighter
  Mamba-MinGRU masker built from Mamba blocks with MinGRU temporal mixing. Replacing the
  ConvTasNet-based auxiliary network with a linear RNN encoder improves both
  signal-to-distortion ratio and predicted MOS while reducing compute. Results establish
  OVC as a practical, low-latency enhancement objective for far-field denoising.
\end{abstract}
%

\section{Introduction}
\label{sec:intro}

The problem of enhancing speech degraded by environmental noise, interfering speakers,
or reverberant effects has been widely studied and remains a common obstacle in many
real-world applications such as telecommunications, smart speakers, and conferencing
devices. A particularly challenging scenario arises when a far-field device, such as a
table-top microphone, captures, enhances, and streams audio back to the user. Because
the acoustic round-trip time through such a pipeline easily exceeds 10\,ms, the user's
own voice arrives with a noticeable delay, producing perceptible echo-like
artifacts~\cite{Stone1999, Groth2004}. Delays beyond 15–20\,ms are widely reported as
disturbing~\cite{Stone2005}, making own-voice suppression an important consideration
for any streamed denoising system operating in far field.

In recent years, deep learning has considerably improved the performance of speech
enhancement models, surpassing classical methods on most publicly available
benchmarks~\cite{Shin2024}. State-of-the-art deep learning-based methods typically come
at the cost of high compute (with associated processing latency), and often require at
least moderate algorithmic latency to achieve strong performance~\cite{Avenstrup2024}.
Linear recurrent models such as Mamba~\cite{Gu2024} and MinGRU~\cite{Feng2024} have
recently emerged as compute-efficient alternatives to transformers that maintain global
temporal context while supporting causal, streaming inference, making them attractive
building blocks for low-latency audio processing.

\begin{figure}[tb]
  \centering
  \includegraphics[width=\columnwidth]{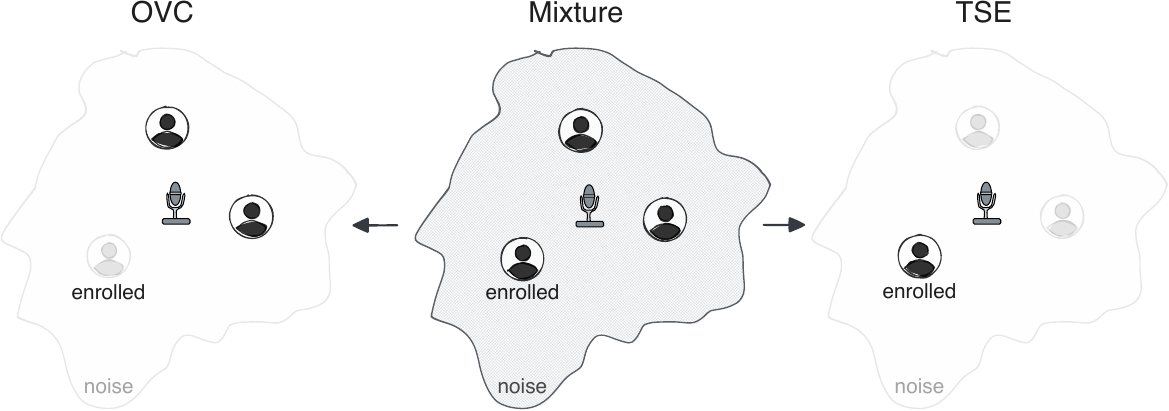}
  \caption{
    Difference between own voice cancellation (OVC) and target speaker
    extraction (TSE). Given a mixture consisting of multiple speakers recorded in a
    noisy scene, TSE (right) aims to keep only the enrolled speaker, while OVC (left)
    removes only the enrolled speaker. Both methods jointly denoise and isolate
    speakers.
  }
  \label{fig:concept}
\end{figure}

In this work we are considering own voice cancellation (OVC), which we define as the
removal of a target speaker from a noisy input mixture of multiple speakers conditioned
on an enrollment utterance from the target speaker. This effectively makes OVC a
generalization of speech enhancement and closely related to target speaker extraction
(TSE), see \cref{fig:concept}.

Our main contributions are as follows:
\begin{enumerate}[label=(\roman{*})]
  \item we establish own-voice cancellation (OVC) as a novel approach for mitigating
    latency-induced distortion in streamed far-field denoising by treating the
    user's voice as an unwanted signal to be suppressed;

  \item we introduce a compute-efficient architecture based on linear RNNs
    (Mamba-MinGRU), matching the performance of ConvTasNet-based networks at a
    fraction of the compute while maintaining only 2\,ms algorithmic latency in all
    causal configurations;

  \item we demonstrate that auxiliary linear RNN-based encoders provide better speaker
    representations compared to ConvTasNet-based ones for speaker conditioning.
\end{enumerate}

\section{Related work}
\label{sec:related-work} A substantial body of work has focused on developing
neural network architectures for speech enhancement and separation~\cite{Luo2018,
Luo2019, Luo2020, Shin2024, Saijo2024}. Notable contributions include
TasNet~\cite{Luo2018}
and ConvTasNet~\cite{Luo2019}, with ConvTasNet frequently serving as a
baseline for evaluating new methods~\cite{Shin2024}.

The perceptual impact of processing latency on own-voice perception has been studied
extensively in the hearing-aid literature. Even modest delays of 4--10\,ms can produce
perceptible disturbances~\cite{Groth2004}, and delays exceeding 15\,ms are rated as
unacceptable by most listeners~\cite{Stone2005}, in part due to interference between
the delayed signal and the listener's own direct sound~\cite{Stone1999, Stone2002}.
Although these studies focus on hearing devices, the same perceptual effects arise
whenever enhanced audio is streamed back to the user with non-negligible latency, as in
far-field denoising setups.

The most closely related task to our setup is \emph{target speaker extraction} (TSE),
where the model extracts a single target speaker from a mixture. Prior work includes
SpeakerBeam~\cite{Delcroix2020}, Listen only to me!~\cite{Delcroix2022},
TEA-PSE~\cite{Ju2022}, and more recently SpeakerBeam-SS~\cite{Sato2024} and an SSL
model-based TSE system~\cite{Peng2024}, which all employ an auxiliary embedding network
to condition the main model on a short enrollment utterance. Our proposed own-voice
cancellation task can also be viewed as a special case of general speech separation,
where individual speakers are estimated from a (potentially noisy) mixture, but where
we also jointly solve the source identifiability problem by directly informing the
network of which speaker to remove ahead of time.

Recent work has explored state-space models and linear recurrent networks for audio
tasks. SepMamba~\cite{Avenstrup2024} applied Mamba blocks to speech separation, and
SpeakerBeam-SS~\cite{Sato2024} replaced ConvTasNet temporal convolutions with
S4D~\cite{Gupta2022} layers for real-time target speaker extraction.
MinGRU~\cite{Feng2024} was proposed as a minimal gated recurrence that can be expressed
as a linear recurrence, enabling efficient parallel training while retaining the
streaming capability of classical RNNs. Our Mamba-MinGRU architecture combines the
Mamba block structure with MinGRU as the temporal mixer, yielding a design that is both
compute-efficient and naturally causal.

A related but distinct problem is \emph{acoustic echo cancellation}
(AEC)~\cite{Cutler2024, Chen2023}, which removes a known playback signal that leaks
back into the microphone. AEC relies on access to the far-end reference signal, whereas
OVC operates from only a short enrollment utterance without requiring the playback
signal, making it applicable to scenarios where no such reference is available.

\section{Methods}
\label{sec:methods}

Given an input mixture $\mathbf{y}= \mathbf{x}^{s}+ \sum_{i \neq s}\mathbf{x}^{i} +
\mathbf{n}$ containing a target speaker $s$ (the own-voice), other speaker(s) $i$, and
a noise signal $\mathbf{n}$, the goal is to recover $\bar{\mathbf{y}}= \sum_{i \neq
s}\mathbf{x}^{i}$. Following~\cite{Delcroix2022}, we train the network using at most
one other speaker.

\subsection{Dataset}
We train on a dynamically mixed dataset using LibriSpeech \cite{Panayotov2015} in
WHAM! noise~\cite{Wichern2019}, as in \cite{Sato2024} (although they use DNS4
noise). During training, we use the train-clean-360 LibriSpeech partition for training
and test-clean for reporting metrics.

Input mixtures are created by sampling two distinct speakers from LibriSpeech and a
noise segment from WHAM. From the enrolled (target) speaker we draw two utterances: one
serves as the enrollment reference and the other is used in the mixture. From the other
speaker we draw only a single utterance. The two speech signals are first mixed at a
signal-to-noise ratio (SNR), which is sampled from $[-5, 5]$ dB, after which noise is
added at an SNR sampled from $[0, 25]$ dB~\cite{Sato2024}. During training, we
independently drop (mute) the other speaker with probability $p _{\text{o}}$ and the
enrolled speaker with probability $p_{e}$, following \cite{Delcroix2022}. If the other
speaker is absent, the target output should correspond to silence, and if the enrolled
speaker is absent, the target is the denoised other speaker. We evaluate our models on
the test set on two SNR conditions as in \cite{Sato2024}: (1) uniformly sampled from
$[10, 20]$ dB when evaluating both speakers present (denoted F) and (2) uniformly
sampled from $[0, 10]$ when evaluating denoising (denoted D).

Additionally, we evaluate on multi-speaker scenarios based on
LibriMix~\cite{Cosentino2020}. We generate the multi-speaker mixtures using the mixing
scripts provided by the original LibriMix
authors\footnote{\url{https://github.com/JorisCos/LibriMix}}, but modified to allow
more than two speakers.

\subsection{Model}
\begin{figure}[tb]
  \centering
  \includegraphics[width=\columnwidth]{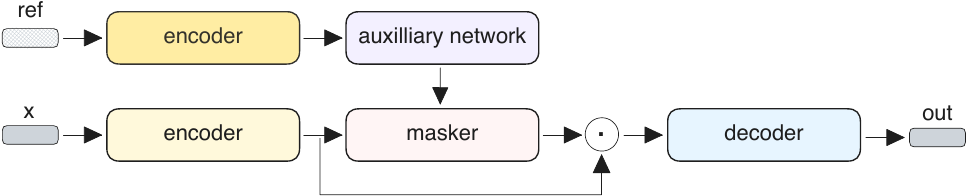}
  \caption{
    High-level architecture of a time-domain conditioned ConvTasNet. Note that
    the two encoders do not share parameters. The output of the auxiliary network
    is an embedding which is applied using an adaptation layer.
  }
  \label{fig:architecture}
\end{figure}

We use TD-SpeakerBeam~\cite{Delcroix2020} as a baseline model and introduce an
alternative, much more compute-efficient, linear RNN-based, architecture. The
high-level architecture is shown in \cref{fig:architecture}.

Our architecture is a time-domain TasNet variant whose masking network is composed
solely of Mamba blocks ~\cite{Gu2024}, using MinGRU~\cite{Feng2024} as the temporal
mixer. We denote this network Mamba-MinGRU.

\begin{figure}[tb]
  \centering
  \includegraphics[width=\columnwidth]{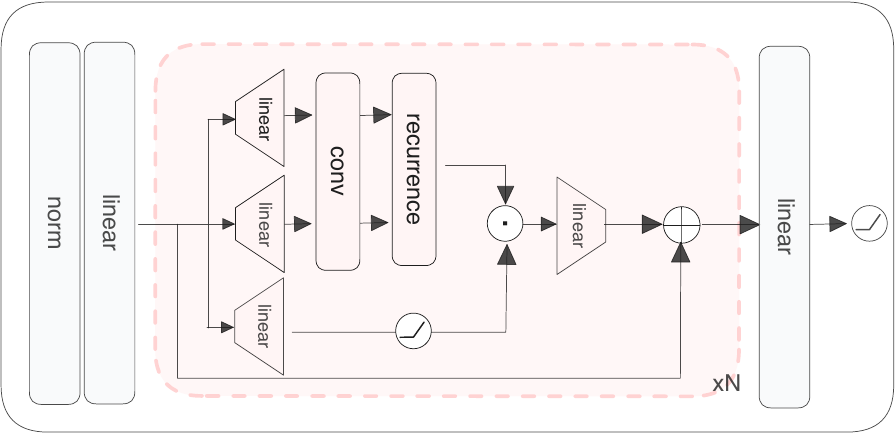}
  \caption{
    Detailed architecture of the Mamba-MinGRU masker. It contains an initial
    normalization layer, followed by a projection to $d_{model}$, and then N
    Mamba-MinGRU
    blocks. A final projection projects the predicted mask back to the encoder
    dimension and applies a non-linearity, here a Sigmoid.
  }
  \label{fig:masker}
\end{figure}

Each Mamba-MinGRU block is a pre-norm residual block consisting of (1)
LayerNorm~\cite{Ba2016}, (2) linear expansion by factor $K$ split into $y,z$, (3) short
causal depthwise 1-D conv + SiLU~\cite{Elfwing2017}, (4) MinGRU recurrence as time
mixing, (5) gating: $y \odot \operatorname{SiLU} (z)$, and finally, (6) linear
projection back to input channels. See \Cref{fig:masker} for a visualization.

The MinGRU recurrence is given by the following equations:
\begin{equation}
  \begin{aligned}
    (\sigma(\mathbf{z}_{t}), \tilde{\mathbf{h}}_{t}) & = \mathrm{split}(y_{t})
    \\
    \mathbf{h}_{t}                                   & = (1 - \mathbf{z}_{t})
    \odot \mathbf{h}_{t-1}+ \mathbf{z}_{t}\odot \tilde{\mathbf{h}}_{t},        \\
  \end{aligned}
\end{equation}
which can be written as a linear recurrence and implemented using a parallel
associative scan~\cite{Orvieto2023} by setting $\text{gates}=( 1-\mathbf{z}_{t})$ and
$\text{tokens}=\mathbf{z}_{t}\odot \tilde{{\mathbf{h}}_t}$,
\begin{equation}
  \mathbf{h}_{t}= \text{gates}\odot\mathbf{h}_{t-1}+ \text{tokens}.
\end{equation}
The recurrence can also be bidirectional, which we implement using Hydra
bidirectionality~\cite{Hwang2024}.

This design yields lower computational complexity than TD-SpeakerBeam while retaining
global context. Regardless of block design, the network consists of an auxiliary
network responsible for extracting speaker embeddings from the enrolled speaker, and a
main network performing own-voice cancellation.

\subsection{Auxiliary network and adaptation}
We investigate two auxiliary networks: (1) a ConvTasNet network with a single
repetition as in \cite{Delcroix2022} and (2) a bidirectional linear RNN network
using only 5 blocks.

Adaptation refers to how the main network is conditioned on the speaker embedding. We
use element-wise multiplication of the embedding and the intermediate representation.
When the network uses skip connections, the auxiliary network outputs one embedding for
the skip path and one for the residual~\cite{Delcroix2020}.

\subsection{Loss function}
The networks are optimized using negative thresholded signal-distortion ratio (SDR)
loss~\cite{Wisdom2020}, but extended to handle silence~\cite{Delcroix2022}:
\begin{equation}
  \mathbf{L}_{\text{SDR}}\left(\hat{\mathbf{x}}, \mathbf{x}, \mathbf{y}\right) =
  \begin{cases}
    \mathcal{L}^{\text{active}}(\mathbf{\hat{x}}, \mathbf{x}),   & \text{if
    $\mathbf{x}$ $\neq \mathbf{0}$},                                        \\
    \mathcal{L}^{\text{inactive}}(\mathbf{\hat{x}}, \mathbf{y}), & \text{if
    $\mathbf{x}= \mathbf{0}$},
  \end{cases}
\end{equation}
The active case is used when there is a speaker present apart from the enrolled speaker,
and the inactive case when there is only the enrolled speaker, in which case the network
should predict silence:
\begin{equation}
  \mathcal{L}^{\text{active}}(\hat{\mathbf{x}}, \mathbf{x}) = - 10 \log_{10}\left
  ( \frac{||\mathbf{x}||^{2}}{||\mathbf{x}- \hat{\mathbf{x}}||^{2}+ \tau
    ||\mathbf{x}||^{2}}
  \right),
\end{equation}
\begin{equation}
  \mathcal{L}^{\text{inactive}}(\hat{\mathbf{x}}, \mathbf{y}) = 10 \log_{10}\left
  ( ||\hat{\mathbf{x}}||^{2}+ \tau ||\mathbf{y}||^{2}\right),
\end{equation}
with $\tau$ being a soft threshold. We set $\tau$ to $10^{-3}$ and $10^{-2}$ for the
active and inactive losses, respectively. The soft thresholds ensure that the model
does not continue to improve on a mixture that is already well separated.

\subsection{Evaluation metrics}
We report two complementary metrics: SDR~\cite{Roux2018} and predicted mean
opinion score (PMOS). As MOS predictor, we use DistillMOS~\cite{Stahl2025} to
get a predicted MOS for each enhanced waveform and average it across utterances.
We report both metrics on two conditions: (F) full mixtures (own voice present) and (D)
denoising-only mixtures (own voice absent).

\subsection{Experimental settings}
For all our experiments we use a sample rate of 16 kHz. Each batch consists of an
enrollment utterance with a duration of 2 seconds and an input mixture of 3 seconds.
We train each model for 1 million steps with a batch size of 8, using AdamW as
optimizer~\cite{Loshchilov2019}
combined with a simple linear decay-to-zero learning rate schedule~\cite{Bergsma2025}.
We set the initial learning rate to $5 \times 10^{-4}$ for all experiments, and $p
_{\text{o}}$ and $p_{e}$ to $10\%$.

For the baseline TD-SpeakerBeam we use the following hyperparameters (terminology from
\cite{Luo2019}): $N=256$, $L=32$, $B=256$, $H=512$, $P=3$, $X=8$, $R=4$. A kernel size
($L$) of 32 corresponds to an algorithmic delay of 2\,ms at 16\,kHz, which all
configurations share.

For the Mamba-MinGRU network, we set the expansion factor $K$ to 2.0, and the model
dimension to 192 and 128 for the "base" and "small" configurations respectively. Both
configurations use 15 blocks with the adaptation layer placed after the 8th block and
share encoder-decoder configuration with the TD-SpeakerBeam network.

Real-time factor (RTF) is measured in causal streaming mode on an Intel Core i7-13700
CPU, processing one block of 16 samples (1\,ms at 16\,kHz) at a time with a single
thread. We export the model using ExecuTorch and perform the RTF measurements using
the C++ runtime. We report the median RTF over 1000 forward passes, discarding the
first 5 as warmup.

\section{Results and discussion}
\label{sec:results}

\begin{table*}[!tb]
    \centering
    \caption{Evaluation results on the dynamic OVC test set. Causal $\checkmark$ indicates
    causal (streaming) inference. OVC: own-voice cancellation; TSE: target speaker extraction; F: full mixture (enrolled speaker present);
    D: denoising only (enrolled speaker absent); SDR: signal-to-distortion ratio
    improvement; pMOS: predicted mean opinion score; MACs: multiply-accumulate operations; RTF: real-time factor; aux: auxillary (speaker embedding) network. The best performing model in~F is marked with bold, while the best performing causal model is indicated with underline.}
    \label{tab:main-results}
    \begin{tabular}{@{}rlccc|ccrr|rrrr@{}}
        \toprule
        & \multirow{2}{*}{Method} & \multirow{2}{*}{Task} & \multirow{2}{*}{Causal} & \multirow{2}{*}{RTF} & \multicolumn{2}{c}{SDR (dB)} & \multicolumn{2}{c|}{pMOS} & \multicolumn{2}{c}{\shortstack{Params (M)}} & \multicolumn{2}{c}{\shortstack{MACs (G/s)}} \\
        \cmidrule(lr){6-7} \cmidrule(lr){8-9} \cmidrule(lr){10-11} \cmidrule(lr){12-13}
        & & & & & F & D & F & D & main & aux & main & aux \\
        \midrule
         & Mixture & - & - & - & -0.07 & 5.02 & 3.28 & 2.95 & - & - & - & - \\
        \midrule
        (a1) & TD-SpeakerBeam & TSE &  &  & 13.66 & 1.14 & 3.15 & 1.55 & 4.94 & 1.66 & 4.97 & 1.67 \\
        (a2) & TD-SpeakerBeam & TSE & $\checkmark$ &  & 11.01 & 9.18 & 2.56 & 2.30 & 4.94 & 1.66 & 4.94 & 1.67 \\
        \midrule
        (b1) & TD-SpeakerBeam & OVC &  &  & 13.42 & 14.78 & 3.19 & 3.26 & 4.94 & 1.66 & 4.97 & 1.67 \\
        (b2) & TD-SpeakerBeam & OVC & $\checkmark$ &  & 11.13 & 12.09 & 2.66 & 2.64 & 4.94 & 1.66 & 4.94 & 1.67 \\
        \midrule
        (c1) & Linear RNN & OVC &  &  & 13.38 & 14.93 & 3.22 & 3.32 & 4.71 & 1.65 & 0.33 & 1.67 \\
        (c2) & \quad + Linear RNN emb. & OVC &  &  & \textbf{13.57} & 9.67 & 3.20 & 2.71 & 4.71 & 1.61 & 0.33 & 0.26 \\
        (c3) & Linear RNN & OVC & $\checkmark$ & 1.69 & 11.50 & 12.46 & 2.76 & 2.71 & 4.72 & 1.65 & 0.33 & 1.67 \\
        (c4) & \quad + Linear RNN emb. & OVC & $\checkmark$ & 1.69 & \underline{11.98} & 11.35 & 2.80 & 2.65 & 4.72 & 1.61 & 0.33 & 0.26 \\
        (d1) & Linear RNN (small) & OVC & $\checkmark$ & 0.82 & 11.21 & 12.33 & 2.66 & 2.63 & 2.17 & 1.65 & 0.18 & 1.66 \\
        (d2) & \quad + Linear RNN emb. & OVC & $\checkmark$ & 0.82 & 11.47 & 11.25 & 2.71 & 2.55 & 2.17 & 1.63 & 0.18 & 0.26 \\
        \bottomrule
    \end{tabular}
\end{table*}

Our main results are shown in \cref{tab:main-results}. When comparing task objectives,
OVC and TSE appear comparably difficult (cf.\ (a1) vs.\ (b1)), with both achieving
$\sim$13\,dB SDR in the full mixture condition~(F). Moving to a causal setting incurs a
moderate drop in SDR for both tasks (cf.\ (a1) vs.\ (a2) and (b1) vs.\ (b2)).

Replacing the TD-SpeakerBeam masking network with the proposed Mamba-MinGRU
architecture (c1) yields competitive non-causal performance at a fraction of the
compute: the main network uses only 0.33\,GMAC/s compared to 4.97\,GMAC/s for
TD-SpeakerBeam. In the causal setting, the linear RNN model~(c3) closely matches the
causal TD-SpeakerBeam baseline~(b2) while remaining far more efficient.

Substituting the ConvTasNet-based auxiliary encoder with a linear RNN encoder further
reduces auxiliary compute from 1.67\,GMAC/s to 0.26\,GMAC/s while improving SDR on the
full mixture condition in all settings. In the non-causal setting, the linear RNN
auxiliary encoder~(c2) improves SDR on the full mixture condition to 13.57\,dB. In the
causal setting~(c4), SDR is comparable to or better than the ConvTasNet-based auxiliary
while using substantially less compute. Using a linear RNN auxiliary encoder does seem
to trade better performance in~F for a drop in performance in denoising~(D). This trend
becomes increasingly more apparent when the main network becomes more expressive,
see~(c2) versus~(c1).

The small Mamba-MinGRU variant~(d1/d2), with roughly half the main-network parameters
of the full model, still achieves competitive causal performance (11.47\,dB SDR on~F)
at 2\,ms algorithmic latency, demonstrating the scalability of the approach to
resource-constrained streaming devices.

Although the base model~(c4) achieves an RTF of 1.69, the smaller variant~(d2) runs
below real-time with an RTF of 0.82 using a single CPU thread. While further hardware
optimizations should be considered especially with regards to the larger model, this
demonstrates that the compact variant is already suitable for real-time streaming with
very low (2\,ms) latency. This should be compared against the reported Speakerbeam-SS
results, which show an RTF below 1 but with an algorithmic latency of 20\,ms~\cite{Sato2024}, which is
above the perceptual threshold for own-voice artifacts.

\subsection{Importance of speaker pitch}

We show the effect of speaker pitch, as determined by the fundamental frequency $\left
(f_{0}\right)$ of the enrollment and target speakers, in~\cref{tab:pitch-res}. For each
subject in the test set, we compute a subject-specific $f_{0}$ using
PYIN~\cite{Mauch2014}, which is a probabilistic version of the YIN pitch tracking
algorithm~\cite{DeCheveigne2002}, as the mean $f_{0}$ across all segments of speech.
These subject-specific $f_{0}$ scores are then stratified into high and low pitch bins
with a cutoff at 160 Hz, and we report model performance on these splits.

We observe that it is more difficult to remove own voice speech from mixtures wherein
both speakers have the same pitch compared to when the speakers have different $f
_{0}$. It appears to be slightly easier to remove own voice when the enrolled speaker
has low $f_{0}$ (cf. f1 vs. f2).

\begin{table}[t]
	\centering
	\small
	\caption{%
	Importance of differences in fundamental pitch frequencies for high ($>$160 Hz)
	and low ($<$160 Hz) pitch groups evaluated on model (c3).
	}
	\label{tab:pitch-res}
	\setlength{\tabcolsep}{3pt}
	\begin{tabular}{@{}llc@{}}
		\toprule      & Method                             & SDR   \\
		\midrule (e1) & Same pitch - high                  & 11.09 \\
		(e2)          & Same pitch - low                   & 10.91 \\
		(f1)          & Different pitch - enrolled is high & 11.45 \\
		(f2)          & Different pitch - enrolled is low  & 12.24 \\
		\bottomrule
	\end{tabular}
\end{table}

\subsection{Multiple speakers}
\begin{figure}[tb]
  \centering
  \includegraphics[width=\columnwidth]{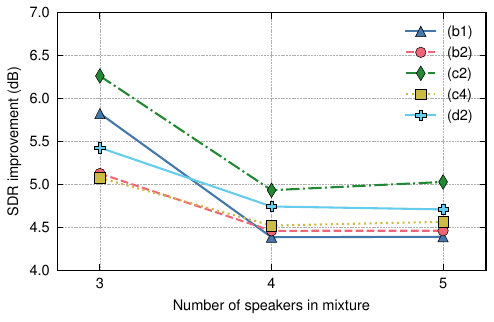}
  \caption{
    SDR improvement (dB) for mixtures with multiple interfering speakers. Model
    IDs in the legend correspond to those shown in~\cref{tab:main-results}.
  }
  \label{fig:multi-speaker}
\end{figure}

We evaluate the robustness of our OVC models to scenarios with mixtures with 3,
4, and 5 speakers in~\cref{fig:multi-speaker}. For all three evaluated models,
performance
in SDR improvement degrades significantly with a general drop of $\sim$2 dB with
increasing number of mixture speakers. This degradation is expected as the acoustic
scene becomes increasingly complex, making it more challenging to isolate and
suppress the enrolled speaker's voice.
We leave it as future work to extend this method to more than two speakers.

\section{Conclusion}
\label{sec:conclusion}

We have introduced own-voice cancellation as a practical objective for far-field streamed
denoising, showing that methods from target speaker extraction can effectively
remove an enrolled speaker from a noisy mixture. The proposed Mamba-MinGRU architecture
matches the performance of ConvTasNet-based baselines at a fraction of the compute,
and replacing the auxiliary encoder with a linear RNN further reduces cost while
maintaining competitive performance. With only 2\,ms algorithmic latency in all causal
configurations, these results pave the road for low-footprint own-voice
suppression in streaming devices. Future work includes training with more than two
simultaneous speakers and evaluating under reverberant conditions.

\section{Generative AI Use Disclosure}
\label{sec:generative_ai_use_disclosure}

In accordance with ISCA policy, generative AI tools were not used as co-authors, nor to develop the source code.
AI tools were used solely for grammar correction.

\bibliographystyle{IEEEtran}
\bibliography{references}

\end{document}